\documentclass[reprint, amsmath,amssymb,aps,letter]{revtex4-1}

\usepackage{float}
\usepackage{graphicx}% Include figure files
\usepackage{dcolumn}% Align table columns on decimal point
\usepackage{bm}% bold math

\usepackage[usenames,dvipsnames]{color}

\usepackage{longtable}
\usepackage{multirow}
\usepackage{comment}
\usepackage{soul}
\usepackage[normalem]{ulem}

\begin{document}

\preprint{APS/123-QED}

\title{Probing the Viscoelastic Properties of Aqueous Protein Solutions using Molecular Dynamics Simulations}% Force line breaks with \\

\author{D. F. Hanlon}
\email{dfh031@mun.ca}
\author{Ivan Saika-Voivod}
\author{G. Todd Andrews}
\affiliation{Department of Physics and Physical Oceanography, Memorial University of Newfoundland, St. John's, NL, Canada, A1B 3X7}

\author{M. Shajahan G. Razul}
\affiliation{Department of Chemistry, St. Francis Xavier University, Antigonish, NS, B2G 2W5, Canada}

\date{\today}% It is always \today, today,
\begin{abstract}
We performed molecular dynamics simulations to investigate the viscoelastic properties of aqueous protein solutions containing an antifreeze protein, a toxin protein, and bovine serum albumin. These simulations covered a temperature range from 280 K to 340 K. Our findings demonstrate that lower temperatures are associated with higher viscosity as well as a lower bulk modulus and speed of sound for all the systems studied. Furthermore, we observe an increase in the bulk modulus and speed of sound as the temperature increases up to a weak maximum while the viscosity decreases. Moreover, we analyzed the influence of protein concentration on the viscoelastic properties of the antifreeze protein solution. We observed a consistent increase in the bulk modulus, speed of sound, and viscosity as the protein concentration increased. Remarkably, our molecular dynamics simulations results closely resemble the trends observed in Brillouin scattering experiments on aqueous protein solutions.  The similarity thus validates the use of simulations in studying the viscoelastic properties of protein water solutions. Ultimately, this work provides motivation for the integration of computer simulations with experimental data and holds potential for advancing our understanding of both simple and complex systems.

% \begin{description}
% \item[PACS numbers]
% May be entered using the \verb+\pacs{#1}+ command.
 
% \end{description}
\end{abstract}

\pacs{Valid PACS appear here}% PACS, the Physics and Astronomy
\maketitle

\section{Introduction} 

Water is vital for life on Earth yet there are still many properties of water and its interaction with other molecules that are not fully understood.  In particular, there has been considerable recent interest in the viscoelastic properties of aqueous biomacromolecular solutions, especially water-protein solutions, due to their importance in understanding fundamental biophysical and biochemical processes.    

Various experimental techniques have been used to study the viscoelastic properties of aqueous protein solutions  \cite{hanlon2023temperature, denn1984, sharma2011rheology,dmitriev2019phase}. A previous Brillouin scattering study examined aqueous solutions of lysozyme and bovine serum albumin (BSA) with varying concentrations \cite{dmitriev2019phase}. The study found that increasing protein concentration increases both sound velocity and viscosity, determined by spectral peak position and peak full width at half maximum (FWHM), respectively. In a rheology study, the bulk and viscoelastic properties of BSA at low concentrations were investigated \cite{sharma2011rheology}. Both the Brillouin and rheology studies showed a strong dependence of viscoelasticity on protein concentration. Similarly, another study used Brillouin light scattering spectroscopy to examine the temperature dependence of the viscoelastic properties of snail mucus, a natural system consisting primarily of water and glycoprotein  \cite{hanlon2023temperature}. As the temperature increased, the speed of sound and storage modulus increased, while viscosity and hypersound attenuation decreased. Another notable feature present in the mucus was a liquid-to-solid phase transition at around 270.5 K. Further experiments on diluted and dehydrated snail mucus showed significant variations in speed of sound, viscosity, and sound absorption as a function of protein concentration \cite{hanlon2023influence}. Rheology studies were also conducted on dehydrating snail mucus and revealed that the shear modulus increased with increasing degree of dehydration \cite{denn1984}. Overall, previous experimental studies have consistently demonstrated that the viscoelasticity of aqueous protein solutions is strongly influenced by protein concentration \cite{hanlon2023temperature, denn1984, sharma2011rheology,dmitriev2019phase}.

Somewhat surprisingly, molecular dynamics (MD) simulations have not been used much for investigating bulk viscoelastic properties such as bulk modulus, speed of sound, and viscosity of aqueous protein solutions nor fluids in general. Only a few MD studies have explored the dynamical properties of aqueous protein solutions containing collagen molecules and different co-solvents \cite{gautieri2012viscoelastic,wei2010effects}. Specifically, MD simulations have been employed to determine the viscosity and relaxation times of collagen molecules, aiming to explore their elastic behavior and investigate the possible role of cross-linked collagen molecules in viscoelastic properties, as previously hypothesized \cite{gautieri2012viscoelastic, silver2001molecular, silver2001transition}. A likely reason for the limited studies on the viscoelastic properties of aqueous protein solutions is possibly due to the overall accuracy and reliability of the simulations which depend on a number of factors, including the choice of force field, system size, and simulation time.
% \subsection{Concise Overview of Current Work and Significance}

In this paper we demonstrate the capabilities of MD simulations in exploring fundamental properties of fluids by obtaining bulk viscoelastic properties of protein-water systems over the temperature range 280 K $\leq T \leq$ 340 K.  We found that the magnitude and temperature dependence of the bulk modulus, sound speed, and shear viscosity of simulated aqueous protein solutions containing an antifreeze protein, a toxin protein, and BSA are similar to those observed in previous experimental studies on aqueous protein solutions.  The combination of MD simulations with experimental results therefore provides a more comprehensive understanding of the physics of molecular interaction in water-macromolecule solutions that could lead to advancements in fields such as drug discovery, materials science, biochemistry and biophysics. 

\section{Computational Details}
Table \ref{tab:systems} gives an overview of the simulated systems studied in the present work. These included aqueous solutions of an antifreeze glycoprotein found in eelpout, a cross-linking toxin protein found in \textit{Vibrio cholerae}, and the widely studied bovine serum albumin \cite{jia1996structural,durand2012crystal,bujacz2012structures}. The initial protein configurations were obtained from the Protein Data Bank (PDB).  The naming convention employed to designate each system consists of the PDB ID of the protein followed by the weight percentage of that protein in the system. For instance, antifreeze protein 1MSI at 4 wt\% is referred to as 1MSI-4. The antifreeze glycoprotein 1MSI was used due to its simplistic structure and the fact that antifreeze glycoprotein solutions have recently been the subject of experimental work \cite{hanlon2023influence,hanlon2023temperature}. The toxin protein with PDB ID 4GQK was used primarily for its ability to cross-link.  Lastly, BSA (PDB ID 4F5S) was used since it is an extensively studied molecule. Moreover, in addition to the specific reasons for using the chosen proteins, their collective inclusion enables a comprehensive study of proteins characterized by progressively higher molecular weights. This allows for the potential to provide insights into the influence of protein size on the viscoelastic properties of such systems.

\begin{table*}[t]
% \scriptsize
  \centering
  \caption{Overview of protein-water systems used in the present molecular dynamics simulations.  The temperature range studied was 280 K - 340 K.}
  \label{tab:systems}
  \begin{tabular}{ccccc}
    \hline\hline
    System $\rightarrow$  & 1MSI/Water\footnotemark[1] & 4GQK/Water  & 4F5S/Water   \\
    \hline
    \multirow{2}{*}{Protein Data Bank ID} & \multirow{2}{*}{1MSI} & \multirow{2}{*}{4GQK} & \multirow{2}{*}{4F5S}  \\
    & & & \\
    
    \multirow{2}{*}{Protein } & \multirow{2}{*}{\parbox[c]{4cm}{Antifreeze Glycoprotein QAE(HPLC 12)}} & \multirow{2}{*}{\parbox[c]{3cm}{VgrG1-ACD with ADP}} & \multirow{2}{*}{\parbox[c]{3cm}{Bovine Serum Albumin}}  \\
    & & & \\

    \multirow{2}{*}{Weight of Protein (kDa)} & \multirow{2}{*}{7.4} & \multirow{2}{*}{44.3} & \multirow{2}{*}{133.3}   \\
    & & &\\

    \multirow{2}{*}{\# of Protein Molecules in System} & \multirow{2}{*}{1, 3, 5} & \multirow{2}{*}{2} & \multirow{2}{*}{1} \\
    & & &  \\

    \multirow{2}{*}{\# of Water Molecules in System} & \multirow{2}{*}{9970, 8110, 7190} & \multirow{2}{*}{52,030} & \multirow{2}{*}{258,840} \\
    & & &  \\

    \multirow{2}{*}{Protein Concentration (wt \%)} & \multirow{2}{*}{4, 12, 22} & \multirow{2}{*}{4} & \multirow{2}{*}{3} \\
    & & &  \\

    \multirow{2}{*}{System Volume (nm$^{3}$)} & \multirow{2}{*}{6.5$\times$6.5$\times$6.5} & \multirow{2}{*}{12$\times$12$\times$12} & \multirow{2}{*}{20$\times$20$\times$20} \\
    & & &  \\

    \multirow{2}{*}{System Density\footnotemark[2] (g/cm$^3$)} & \multirow{2}{*}{ 1.009, 1.025, 1.060} & \multirow{2}{*}{1.024} & \multirow{2}{*}{1.010} \\
    & & &  \\

    \multirow{3}{*}{Rationale for Use} & \multirow{2}{*}{\parbox[c]{3cm}{Simple structure \& an antifreeze glycoprotein}} & \multirow{2}{*}{\parbox[c]{3cm}{Ability to cross-link in solution}} & \multirow{2}{*}{\parbox[c]{3cm}{Availability \& a widely studied protein}} \\
    & & &  \\
    & & & \\
    & & & \\

    \multirow{2}{*}{Miscellaneous Notes} & \multirow{2}{*}{\parbox[c]{3cm}{Antifreeze glycoprotein found in eelpout \cite{jia1996structural}}} & \multirow{2}{*}{\parbox[c]{3cm}{Toxin protein
found in \textit{Vibrio cholerae} \cite{durand2012crystal}  }} & \multirow{2}{*}{\parbox[c]{3cm}{Structure determined from Ref. \cite{bujacz2012structures}}} \\
    & & &  \\
    & & & \\
    \hline \hline
  \end{tabular}
  \footnotetext[1]{System naming convention: Protein Data Bank ID followed by protein concentration in wt\% ({\it{e.g.}}, 1MSI-4 is protein 1MSI at 4 wt\%).}
  \footnotetext[2]{Density of each system at 300 K.}
\end{table*}

The MD simulations were carried out in the isothermal-isobaric (NPT) ensemble using GROMACS v2022.3 \cite{hess2008gromacs,lindahl2001gromacs,van2005gromacs}. We used the TIP4P/2005 water model (volume $4\times4\times4$ nm$^{3}$, density 0.992 g/cm$^{3}$) as a control system due to its ability to accurately represent the properties of water over the temperature range studied \cite{abascal2005general,gonzalez2016comprehensive, yu2023unified}. Proteins in the system interact using the All-Atom Optimized Potential for Liquid Simulations (OPLS-AA) \cite{JorgensenWilliamL1996DaTo} and was chosen due to its good description of liquid organic systems and ability to reproduce thermodynamic properties such as density and heat of vaporization, structural properties, as well as free energies of hydration  \cite{zangi2018refinement}. The equations of motion were integrated with a time step of 2 fs using the leap-frog algorithm \cite{cuendet2007calculation}. Temperature was held constant using a Nos\'e-Hoover thermostat with a time constant of 0.1 ps. The pressure was held constant using the Parrinello-Rahman barostat with a time constant of 2 ps. The viscoelastic properties of each system (see Table \ref{tab:systems}) were investigated over the temperature range 280 K to 340 K.  For each system, the size of simulation box and number of solute molecules vary due to the differing sizes of the proteins used and the desire to replicate approximate protein concentrations used in experimental work.

\section{Results \& Discussion}
\subsection{Bulk Modulus}
\begin{figure}[ht]
\includegraphics[width=1\linewidth]{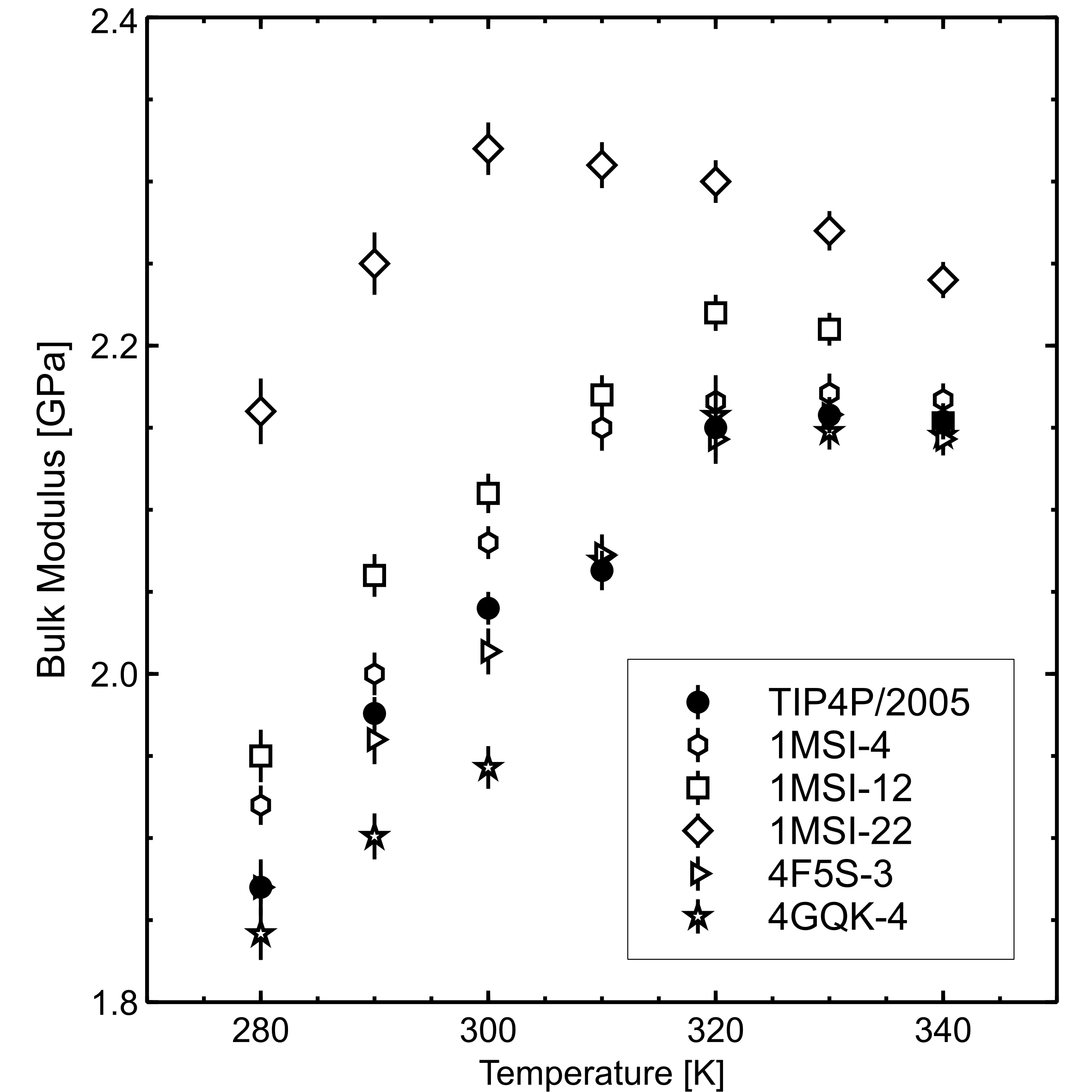}
  \caption{Temperature dependence of bulk modulus for water and aqueous protein solutions with different protein species and concentration.}
  \label{fig:MD_BulkMod}
\end{figure}

Figure \ref{fig:MD_BulkMod} shows the bulk modulus ($K_s$) for solutions studied in this work (see Table \ref{tab:systems}) obtained by analyzing the fluctuations in volume and pressure in the simulations. For most systems, the bulk modulus increases with increasing temperature until it reaches a maximum value near 330 K. However, exceptions to this trend are observed in 1MSI-12 and 1MSI-22, where the bulk modulus exhibits a maximum at 320 K and 300 K, respectively. The difference in the maximum for these two solutions is caused by the increased protein concentration in each system. Additionally, in the case of 1MSI solutions, higher protein concentrations lead to a larger bulk modulus. The presence of a maximum in the bulk modulus also suggests that the molecular structure in each system is most rigid at this temperature.

\begin{figure}[ht]
\includegraphics[width=1\linewidth]{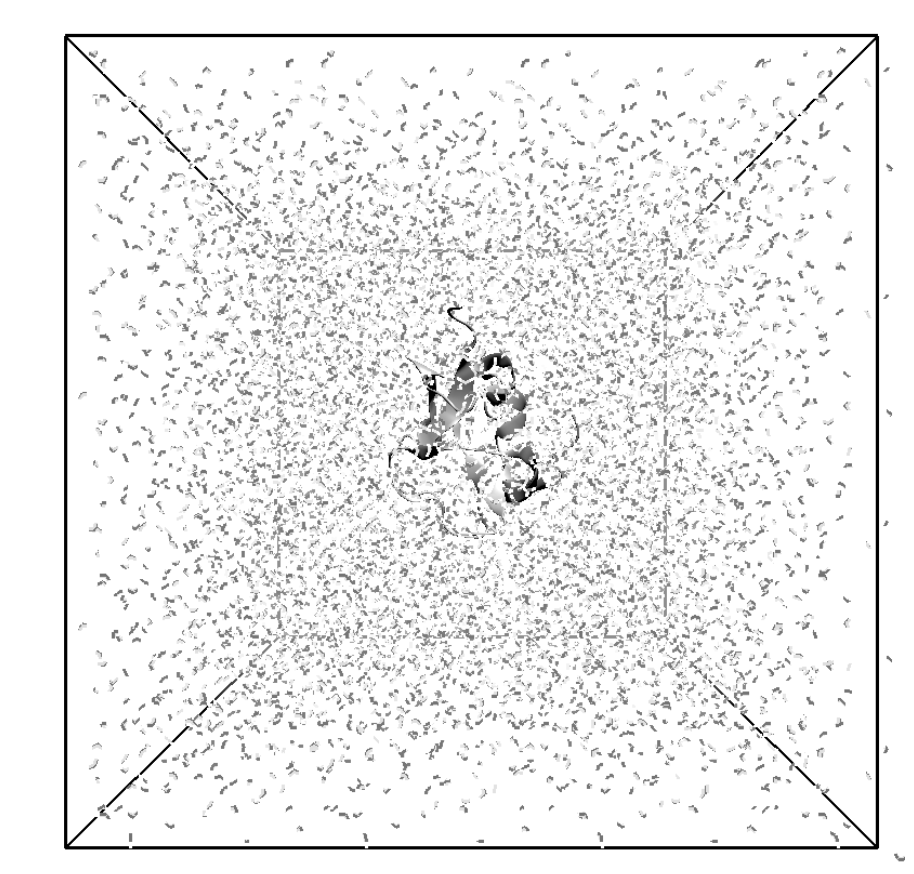}
  \caption{Representative simulation box  of the 1MSI-4 system containing a single 1MSI protein (at center of box) and 2417 water molecules (grey flecks). Figure was produced using VMD v.1.9 \cite{humphrey1996vmd}. }
  \label{fig:1msi_solv}
\end{figure}

\begin{figure}[ht]
\includegraphics[width=1\linewidth]{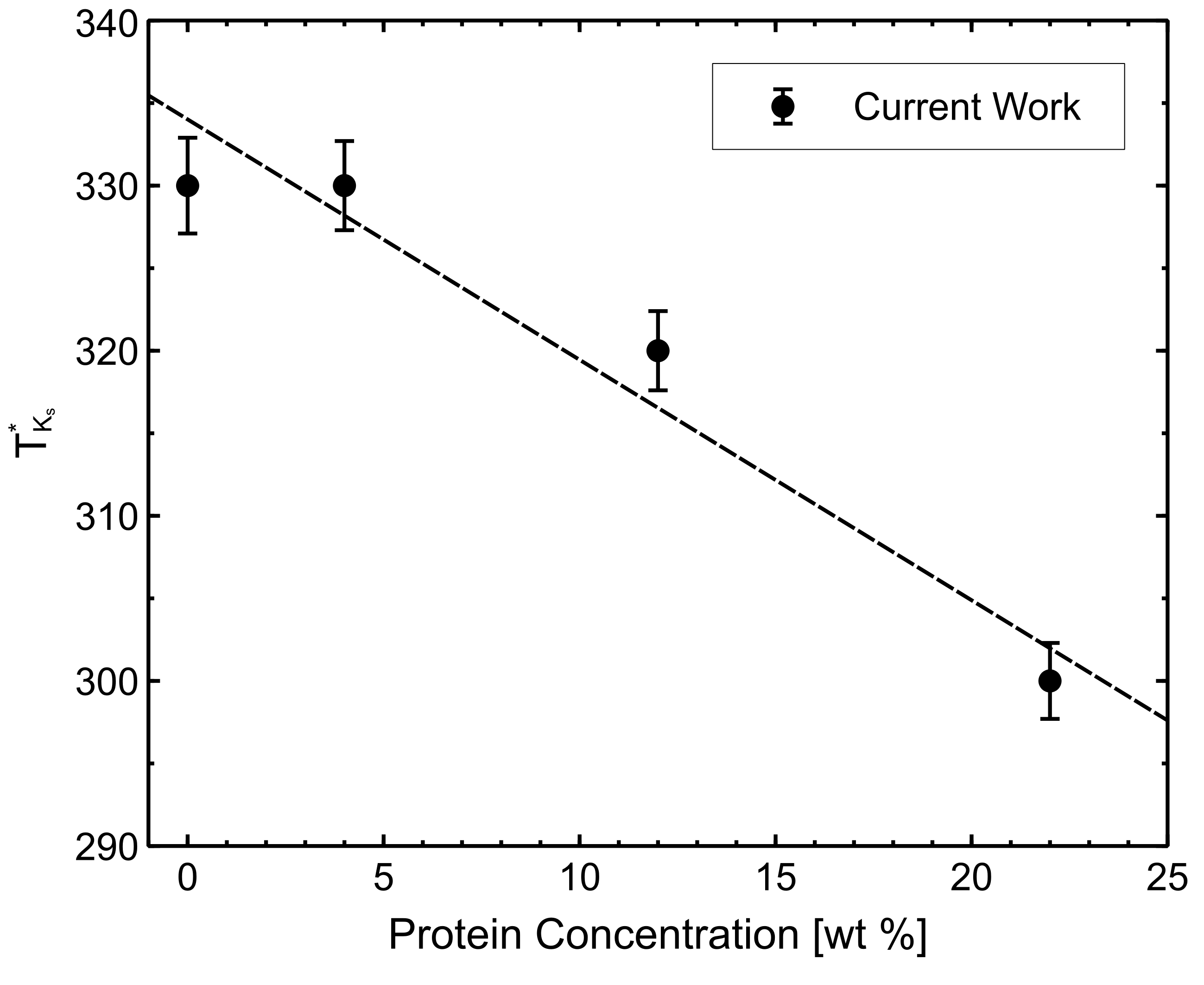}
  \caption{Temperature at which maximum bulk modulus occurs  (T$^*_{K_s}$) as a function of concentration for 1MSI solutions. The line of best fit, represented by the dashed line with the equation T$^*_{K_s}$ = 337[X$_p$] - 1.68, where [X$_p$] denotes the protein concentration. }
  \label{fig:T_ksmax}
\end{figure}

Figure \ref{fig:T_ksmax} illustrates the temperature at which the bulk modulus is at a maximum for each 1MSI solution. It is evident that as protein concentration increases, the temperature at which the maximum bulk modulus occurs decreases. The reason behind this phenomenon is due to the fact that the behavior of water and its bulk modulus (compressibility) is influenced by hydrogen bonds between water molecules and proteins and the presence of proteins has been known to disrupt these bonds \cite{ball2008water}.

\subsection{Sound Velocity}
\begin{figure}[t]
\includegraphics[width=1\linewidth]{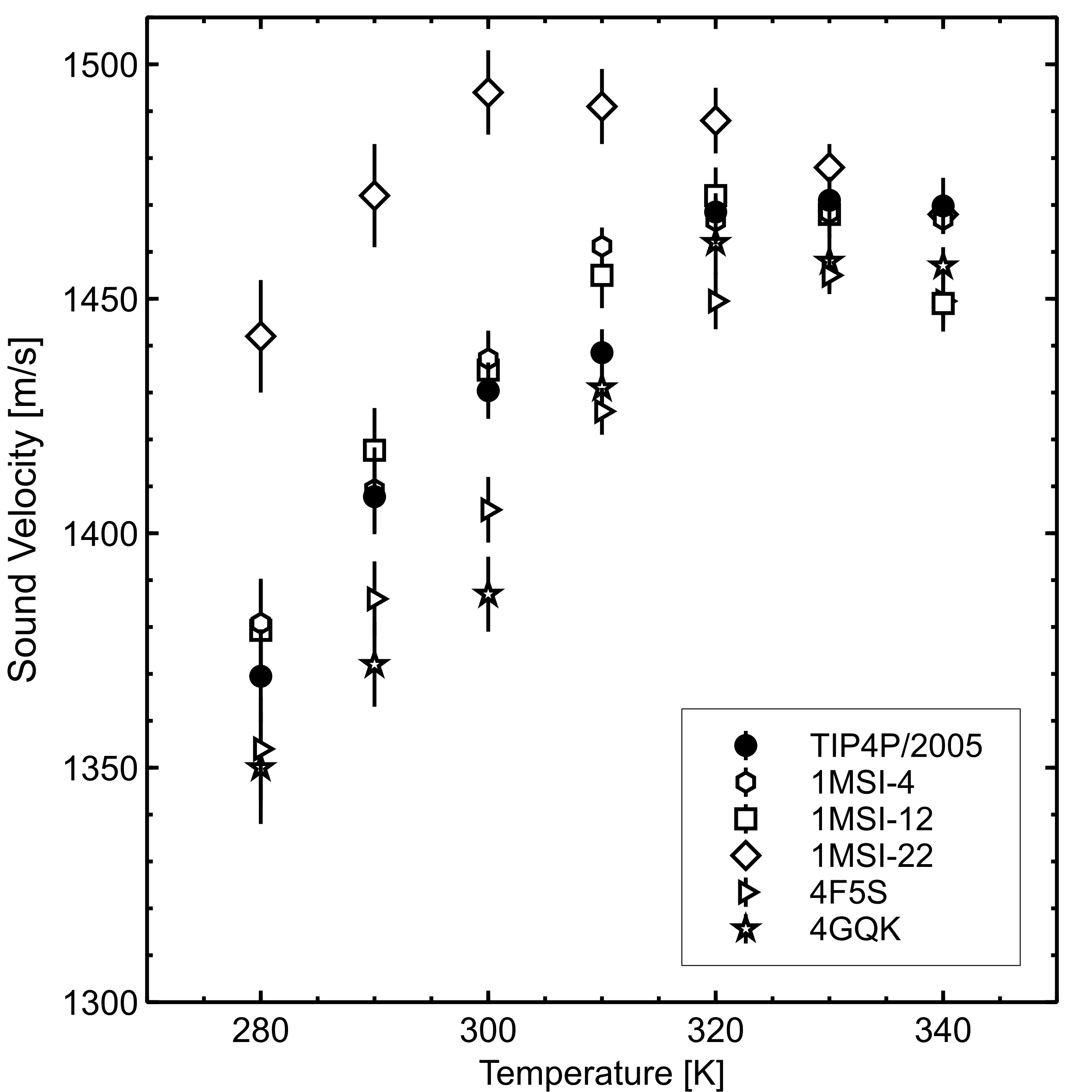}
  \caption{Temperature dependence of sound velocity of water and aqueous protein solutions with different protein species and concentration.}
  \label{fig:MD_SoundVel}
\end{figure}
The temperature dependence of the sound velocity for each of the protein-water systems was determined directly from the bulk modulus and the solution mass density $\rho$ using the relation
\begin{equation}
    v = \sqrt{\frac{K_s}{\rho}}.
    \label{eq:speedofsound}
\end{equation}
As shown in Fig.~\ref{fig:MD_SoundVel}, the behaviour is qualitatively similar for all systems.  There is a steady increase in sound velocity with increasing temperature for temperatures at the lower end of the investigated range while at higher temperatures the velocity decreases with increasing temperature.  The maximum in sound velocity occurs at $\sim330$ K for systems with low protein concentrations (3-4\%) and at slightly lower temperatures for those with higher concentrations (12\% and 22\%).

In the case of the 1MSI systems, the sound velocity displays a strong dependence on concentration. Specifically, an increase in protein concentration generally leads to a higher sound velocity, except for the values at 340 K, where 1MSI-4 exhibits a larger sound velocity than 1MSI-12. The reason for this unexpected behavior at 340 K is currently unknown. Additionally, as the concentration increases, the maximum sound velocity for the 1MSI systems occurs at lower temperatures. This can be seen in Fig. \ref{fig:T_ksmax} which shows the temperature at which the maximum bulk modulus occurs, directly related to the speed of sound via Eq. \ref{eq:speedofsound}. This indicates a complex interplay between concentration, temperature, and sound velocity within the 1MSI system. Notably, except at the highest temperatures, the sound velocities for the 1MSI systems are larger than those for water. Conversely, the 4F5S-3 and 4GQK-4 systems exhibit sound velocity values that closely resemble those of water, although these velocities are generally lower than those of water, despite having protein concentrations similar to 1MSI-4. This observation is interesting because it demonstrates the impact of protein species on sound velocity in aqueous protein solutions, ranging from a decrease relative to v$_{H_2O}$ to an increase relative to v$_{H_2O}$. 

The sound velocities for 1MSI-4 and 1MSI-12 systems are similar for all temperatures but quite different from those of the 1MSI-22 system. Furthermore, the sound velocity for the 1MSI-22 system is also much greater than for all other systems, the difference being most obvious at lower temperatures ($T \leq$ 300 K). This overall larger sound velocity for the 1MSI-22 system can be understood by noting that sound velocity is directly related to bulk modulus.  Therefore, with increased protein concentration, the liquid becomes less compressible than at lower concentrations resulting in an increased bulk modulus and consequently, sound velocity. 

Another notable feature in the bulk modulus and sound velocity data is the near convergence of values for all systems near 340 K (see Fig.~\ref{fig:MD_SoundVel}). At higher temperature this type of feature has been associated with protein denaturation and is consistent with nuclear magnetic resonance studies and MD simulations on lysozyme-water solutions that provide evidence for denaturation at $\sim 340$ K \cite{mallamace2011possible, zhang2009observation}. However, its worth noting that our investigation into the temperature dependent changes in the radius of gyration did not reveal any signs of protein denaturation occurring.

\subsection{Shear Viscosity}

\begin{figure}[t]
\includegraphics[width=1\linewidth]{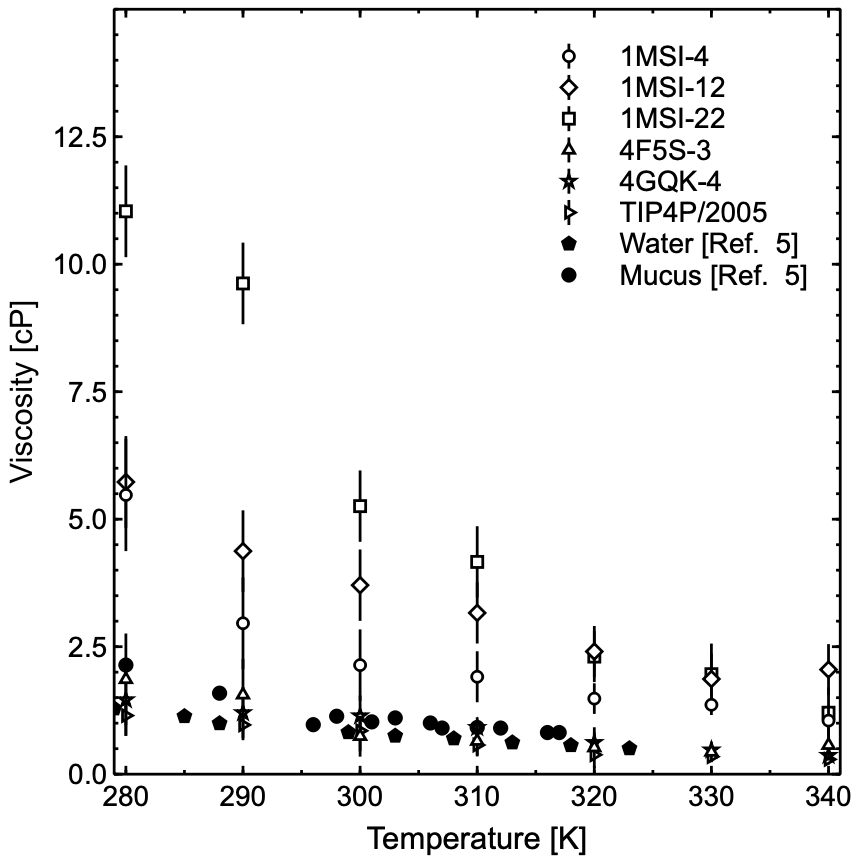}
  \caption{Shear viscosity of water and aqueous protein solutions versus temperature for proteins 4F5S, 4GQK, and 1MSI at concentrations of 4\%, 12\%, and 22\%.  Also shown for the purposes of comparison are apparent viscosity values for water and snail mucus, a natural system consisting primarily of water and glycoproteins, obtained from Brillouin scattering experiments \cite{hanlon2023temperature}.}
  \label{fig:MD_Viscosity}
\end{figure}

Figure \ref{fig:MD_Viscosity} shows the shear viscosity as a function of temperature, calculated using the Einstein relation \cite{hess2008gromacs}
\begin{equation}
    \eta = \lim_{t\to\infty}\frac{V}{k_B T} \frac{d}{dt} \Big \langle \Big ( \int_{t_0}^{t_0 + t} P_{xz}(t^{\prime})dt^{\prime} \Big )^2 \Big \rangle_{t_0},
    \label{eqn:StokeEinstein}
\end{equation}
\noindent 
where $V$ and $T$ are the volume and temperature of the system, respectively.  $P_{xz}(t^\prime)$ is the off-diagonal component of the pressure tensor at time $t^\prime$, $k_B$ is the Boltzmann constant, $t_0$ is the starting time, and $t$ represents the integration time. Our results for the TIP4P/2005 water model show a shear viscosity value of 0.85 $\pm$ 0.09 cP at 300 K, which is in good agreement with the previously obtained value of 0.855 cP at 298 K from Ref. \cite{gonzalez2010shear}. 

Within uncertainty the viscosity of all systems decrease with increasing temperature over the entire temperature range studied. This trend aligns with experimental findings on similar protein-water systems \cite{hanlon2023temperature,lupi2011,comez2012,pochylski2005structural} (see apparent viscosity results for snail mucus, a natural water-glycoprotein system in Fig. \ref{fig:MD_Viscosity}). For the 1MSI solutions, there is an non-linear increase in shear viscosity as the temperature decreases. It is also clear that the shear viscosity of the 1MSI solutions is strongly influenced by protein concentration. In contrast, the magnitude and behaviour of the shear viscosity of the simulated 4GQK-4 and 4F5S-3 solutions are similar to those of TIP4P/2005.

\begin{figure}[t]
\includegraphics[width=1\linewidth]{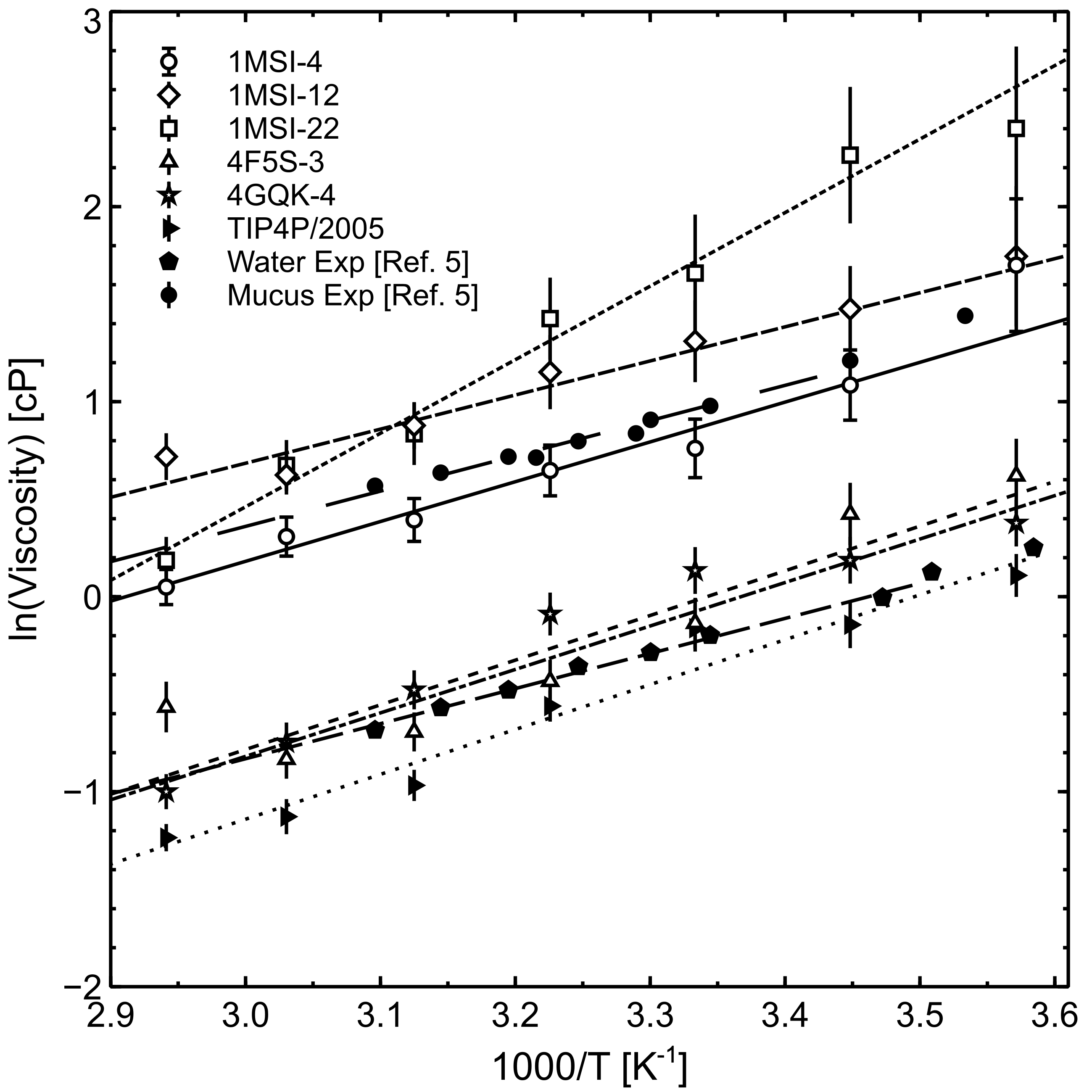}
  \caption{Plot of natural logarithm of shear viscosity of simulated systems as a function of temperature. Solid lines indicate best fits for the entire temperature region. Also shown for the purposes of comparison is the natural logarithm of apparent viscosity for water and snail mucus, a natural system consisting primarily of water and glycoproteins, obtained from Brillouin scattering experiments \cite{hanlon2023temperature}.}
  \label{fig:MD_LogViscosity_2}
\end{figure}

\begin{figure}[t]
\includegraphics[width=1\linewidth]{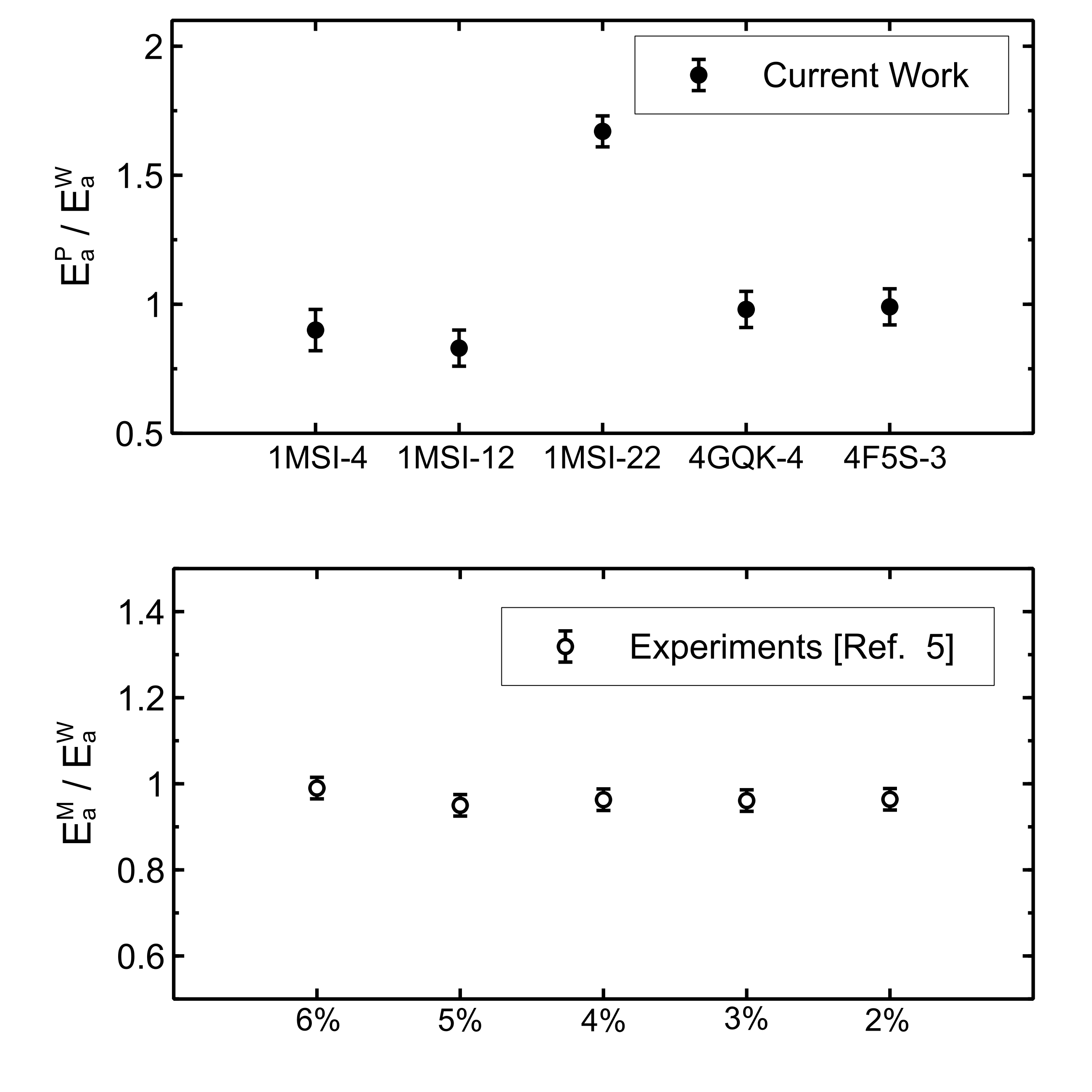}
  \caption{Top Panel: Ratio of activation energy for the simulated aqueous protein solution ($E_a^P$) to activation energy obtained for the TIP4P/2005 water model ($E_a^W$). Bottom Panel: Ratio of activation energy for hydrated snail mucus ($E_a^M$) to activation energy for ultra-pure water ($E_a^W$), as reported in Ref. \cite{hanlon2023influence}.  The labels on the horizontal axis in the lower panel are approximate glycoprotein weight percentages.  Error bars are approximately the size of the symbols.}
  \label{fig:Ea_ratio}
\end{figure}

\begin{table}[!t]
\caption{Best-fit parameters for fit of function $\displaystyle \ln (\eta) = \ln \eta_0 +E_a/k_B T$ to computationally determined shear viscosity.}
\setlength{\extrarowheight}{3pt}
\begin{ruledtabular}
\begin{tabular}{c c c c c}
System & $\ln \eta_0$ & $E_a$ & $R^2$\\
  &(cP) & kJ/mol &  \\
 \hline
1MSI-4& -5.9 $\pm$ 0.8 & 16.7 $\pm$ 0.7 & 0.987\\
1MSI-12& -4.3 $\pm$ 0.9 & 14.5 $\pm$ 0.9 & 0.990\\
1MSI-22& -11.5 $\pm$ 0.9& 31.3 $\pm$ 0.8 & 0.977\\
4GQK-4& -6.3 $\pm$ 0.6 & 18.2 $\pm$ 0.7 & 0.985\\
4F5S-3&  -3.6 $\pm$ 0.9 & 18.5 $\pm$  0.9 & 0.977\\ 
TIP4P/2005&  -7.4 $\pm$ 0.6 & 18.5 $\pm$  0.5 & 0.983\\ \hline
Water\footnotemark[1] & -5.6 $\pm$ 0.3 & 13.8 $\pm$ 0.3& 0.991\\ 
Mucus\footnotemark[2] & -4.3 $\pm$ 0.3 & 13.6 $\pm$ 0.4& 0.988\\ 
Water\footnotemark[3] & --- & 13.6 $\pm$ 0.6 & ---\\ 
    \end{tabular}
    \label{tab:ViscoFit}
    \end{ruledtabular}
    \footnotetext[1]{Water previously studied by Ref.~\cite{hanlon2023temperature}}
    \footnotetext[2]{Mucus previously studied by Ref.~\cite{hanlon2023temperature}}
    \footnotetext[3]{Water previously studied by Ref.~\cite{lupi2011}}
\end{table}
Figure \ref{fig:MD_LogViscosity_2} displays the natural logarithm of shear viscosity as a function of inverse temperature for all systems in this study.  All simulated systems show a linear dependence on $1/T$ over the entire range studied.  We therefore fit the Arrhenius relationship
\begin{equation}
\eta = \eta_0 e^{E_a/k_B T}
\label{eq:lneta}
\end{equation}
\noindent
to the simulation data to obtain the activation energy $E_a$ and entropic pre-factor $\eta_0$ (see Table \ref{tab:ViscoFit}). Notably, with the exception of 1MSI-22 the protein-water systems investigated in the current study have similar values of activation energy, ranging from 16.7 to 18.5 kJ/mol. These findings are consistent with previous experiments on aqueous protein and polymer systems \cite{lupi2011,comez2012,hanlon2023temperature,pochylski2005structural}, where the $E_a$ values were approximately 14-24\% lower than those observed in our simulations. To address this difference we normalized the activation energy relative to ultra-pure water for experimental data and the TIP4P/2005 water model for simulations. This normalization allowed us to mitigate the influence of differences between the ultra-pure water used in experiments and the water model used in simulations. Figure \ref{fig:Ea_ratio} presents the ratios $E^{P}_{a}/E^{W}_{a}$ for the protein-water solutions studied in this work and those obtained from Brillouin spectroscopy for various dilutions of snail mucus  \cite{hanlon2023influence}, $E^{M}_{a}/E^{W}_{a}$. Remarkably, except for 1MSI-22, the ratios for the protein-water systems analyzed in our research were close to unity. This finding aligns with the behavior observed in natural snail mucus, where the activation energy exhibited negligible variation across different concentrations. It is also consistent with the results of a Brillouin scattering study of water-tert-butyl alcohol in which activation energy was found to be nearly independent of polymer concentration \cite{lupi2011}.

As evident from the results displayed in both Fig. \ref{fig:MD_Viscosity} and \ref{fig:MD_LogViscosity_2}, the simulated systems show good agreement with both water and natural snail mucus as previously obtained experimentally \cite{hanlon2023temperature}. For example, considering the 1MSI solutions, which contain antifreeze proteins of varying concentrations, we see very similar behaviour in the viscosity of simulated protein systems with the experimental values obtained for water and natural snail mucus. Additionally, if we consider the data given for the natural logarithm of viscosity as a function of temperature, we see a very similar trend between the 1MSI-4 system and natural snail mucus. In fact, we know from Ref. \cite{hanlon2023temperature} that for T $\leq$ 290 K, the viscous behaviour of snail mucus transition from Arrhenius to non-Arrhenius. We can possibly start to see a similar feature if we consider the single data point at 280 K ($\sim$ 3.6 on x-axis), where the natural logarithm of viscosity increases substantially at this point (see Fig. \ref{fig:MD_LogViscosity_2}). Overall, the behaviour observed in the shear viscosity is consistent with the trends observed in experimental work, providing strong support that the simulations can accurately describe the bulk viscoelastic properties of these systems.

\section{Conclusion}
This study is one of very few, if not only, study to systematically investigate the temperature dependence behavior of viscoelastic properties in aqueous protein solutions using molecular dynamics simulations, while also considering variations in protein sizes and concentration. We determined the temperature dependence of the adiabatic bulk modulus, sound speed, and viscosity, and found behaviour consistent with that observed in previous experimental work on protein-water systems \cite{bail2020,comez2012,lupi2011,pochylski2005structural,hanlon2023temperature, trosel2023diffusion,milewska2003viscosity,hanlon2023influence}.  The general agreement between the simulations results and experimental observations highlights the potential of the former in predicting the viscoelastic properties of macromolecular solutions and in providing complementary molecular-level insight into the behaviour of these systems.

\bibliography{bibliography}

\end{document}